\documentclass[twocolumn]{aastex63}
\usepackage{lineno}
\usepackage{subfigure}
\usepackage{natbib}
\usepackage{graphicx}
\usepackage{float}
\usepackage{hyperref}
\usepackage{xcolor}

\graphicspath{{./}{figures/}}

\shorttitle{Star Formation and AGN Activity in Mrk 709}
\shortauthors{Kimbro et al.}

\begin{document}

\title{Clumpy Star Formation and AGN Activity in the Dwarf-Dwarf Galaxy Merger Mrk~709}

\author{Erin Kimbro}
\affil{eXtreme Gravity Institute, Department of Physics, Montana State University, Bozeman, MT 59717, USA}

\author{Amy E. Reines}
\affil{eXtreme Gravity Institute, Department of Physics, Montana State University, Bozeman, MT 59717, USA}

\author{Mallory Molina} 
\affil{eXtreme Gravity Institute, Department of Physics, Montana State University, Bozeman, MT 59717, USA}

\author{Adam T. Deller}
\affil{Centre for Astrophysics and Supercomputing, Swinburne University of Technology, Hawthorn VIC 3122, Australia}

\and

\author{Daniel Stern}
\affil{Jet Propulsion Laboratory, California Institute of Technology, 4800 Oak Grove Drive, Mail Stop 169-221, Pasadena, CA 91109, USA}

\begin{abstract}

Nearby, low-metallicity dwarf starburst galaxies hosting active galactic nuclei (AGNs) offer the best local analogs to study the early evolution of galaxies and their supermassive black holes (BHs).  Here we present a detailed multi-wavelength investigation of star formation and BH activity in the low-metallicity dwarf-dwarf galaxy merger Mrk 709.  Using {\it Hubble Space Telescope} H$\alpha$ and continuum imaging combined with Keck spectroscopy, we determine that the two dwarf galaxies are likely in the early stages of a merger (i.e., their first pass) and discover a spectacular $\sim 10$ kpc-long string of young massive star clusters ($t \lesssim 10$ Myr; $M_\star \gtrsim 10^5~M_\odot$) between the galaxies triggered by the interaction.  We find that the southern galaxy, Mrk 709 S, is undergoing a clumpy mode of star formation resembling that seen in high-redshift galaxies, with multiple young clusters/clumps having stellar masses between $10^7$ and $10^8~M_\odot$.
Furthermore, we present additional evidence for a low-luminosity AGN in Mrk 709 S (first identified by \citealt{reinesetal2014} using radio and X-ray observations), including the detection of the coronal [Fe X] optical emission line.  The work presented here provides a unique glimpse into processes key to hierarchical galaxy formation and BH growth in the early Universe.

\end{abstract}



\section{Introduction} \label{sec:intro}

Dwarf galaxies form the largest subset of galaxies in the Universe \citep[e.g.,][]{Binggeli}, and their mergers play a crucial role in hierarchical galaxy formation \citep[e.g.,][]{stierwalt2017}. Dwarf-dwarf mergers trigger intense star formation, and lead to the formation of blue compact dwarfs \citep[BCDs;][]{stierwalt2017,Paudel2018,kado-fong2020, zhang2020} that have strongly starbursting spectra and physical sizes of $\lesssim$ 1~kpc \citep{bekki2008}. These low-mass, low-metallicity BCDs can host super star clusters \citep[SSCs;][]{johnson2000,reinesetal2008b}, which have properties expected of the progenitors of globular clusters \citep[e.g.,][]{oconneletal1994}. 

At least some dwarf galaxies with stellar masses comparable to the Magellanic Clouds ($M_\star \sim 10^8-10^{9.5}~M_\odot$) are also known to host massive black holes (BHs) with $M_{\rm BH} \sim 10^{4.5}-10^6~M_\odot$ \citep[see][and references therein]{reines2016,mezcua2017,Greene2020}. These relatively small BHs place our best observational constraints on the masses of the first BH seeds in the earlier Universe. Star-forming dwarfs hosting massive BHs today are particularly interesting (albeit hard to find) since they are more akin to high-redshift galaxies and may help us understand the interplay between BH and galaxy growth at early times.


\begin{figure*}
\centering
\includegraphics[width=0.8\textwidth]{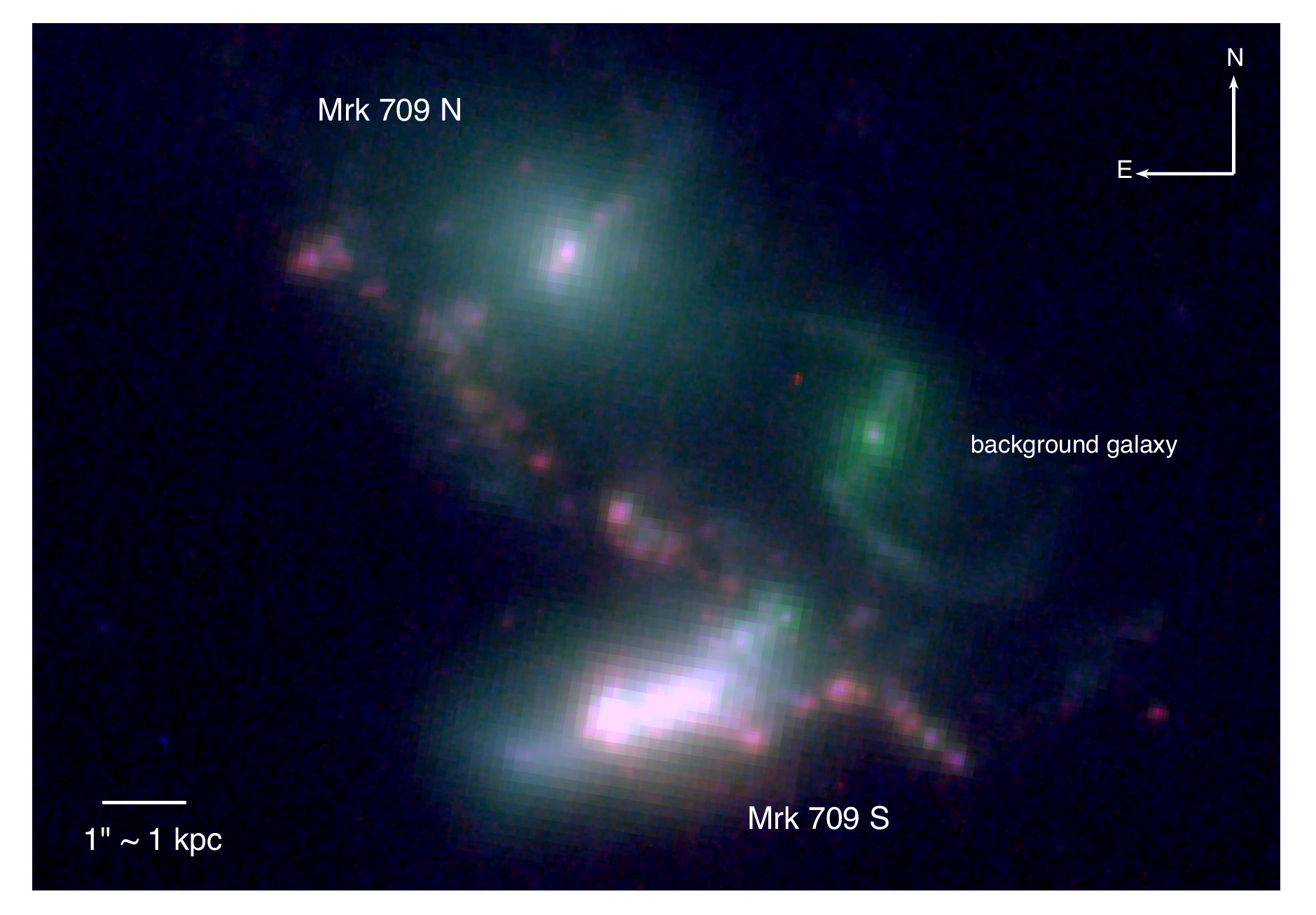}
\caption{
{\it HST}/WFC3 three-color image of Mrk 709. Red shows H$\alpha$ emission (UVIS/F680N), green shows near-infrared continuum (IR/F110W), and blue shows optical continuum (UVIS/F621M). 
The Keck/LRIS 1\farcs0-wide long slit was positioned to cover both the northern and western galaxies, and the redshift of the latter indicated it is a background galaxy and not part of the Mrk 709 system.
The star clusters in the bridge between Mrk 709 N and S are clearly seen in the WFC3 image. 
}\label{fig: Fig 1} 
\end{figure*}

Mrk 709 is uniquely suited for studying both heirarchical galaxy formation and the formation and evolution of BHs. Mrk 709 has been classified as a BCD \citep{gildepaz2003} with very low metallicity \citep[12 + log(O/H) = 7.7, or 10\% $Z_\odot$;][]{masegosa1994}, and \citet{reinesetal2014} find that Mrk 709 consists of a pair of compact dwarf galaxies that appear to be interacting with one another.
Moreover, \citet{reinesetal2014} present observational evidence at radio and X-ray wavelengths for an active massive BH in Mrk 709 S, the southern galaxy in the pair. 
Mrk 709 S has intense, on-going star formation with a star formation rate that is an order of magnitude higher than the Large Magellanic Cloud (and is significantly more luminous than the LMC), yet the two dwarf galaxies have similar stellar masses with $M_\star < 3 \times 10^{9}~M_\odot$ \citep{reinesetal2014,vandermarel2002,Whitneyetal2008}.

In this work, we use multi-wavelength data to confirm the interaction of Mrk 709 N and Mrk 709 S, study star formation throughout the system, and further investigate the active galactic nucleus (AGN) in Mrk 709 S. We use high-resolution H$\alpha$ and continuum imaging from the {\it Hubble Space Telescope (HST)}, Keck optical spectroscopy, and very long baseline interferometry (VLBI) radio imaging with the National Radio Astronomy Observatory (NRAO) High Sensitivity Array (HSA). We also re-analyze the SDSS DR8 spectra of Mrk 709 S to search for the AGN coronal line [Fe X]$\lambda$~6374, to provide further evidence for BH activity. Figure \ref{fig: Fig 1} shows {\it HST} imaging of Mrk 709 and Table \ref{tab:properties} gives basic properties of the two component galaxies, Mrk 709 S and Mrk 709 N.

We describe the observations in Section~\ref{sec:observations} and discuss the Mrk 709 system in Section~\ref{sec:system}. The properties of the star forming regions are presented in Section~\ref{sec:sfreg}. We present and analyze the evidence for the low-luminosity AGN in Mrk 709 S in Section~\ref{sec:agn}, and summarize our results and conclusions in Section~\ref{sec:summary}. In this paper, we assume 
$H_0=73$~km~s$^{-1}$~Mpc$^{-1}$. Mrk 709 S has a redshift of $z$ = 0.052 based on Sloan Digital Sky Survey (SDSS) spectroscopy \citep{reinesetal2014}, which corresponds to a distance $d \sim 214$ Mpc. At this distance, $1\arcsec$ corresponds to $\sim1$~kpc.

\begin{deluxetable*}{cccccccc}
\tablecaption{Galaxy Properties\label{tab:properties}}
\tablewidth{0pt}
\tablehead{
\colhead{Galaxy} & \colhead{RA} & \colhead{Dec} & \colhead{Redshift}
& \colhead{M$_\star$} & \colhead{\it g} & \colhead{\it r} & \colhead{\it i}\\
{} & {(h:m:s)} & {(deg:m:s)} & {} & {($M_\odot$)} & {(mag)} & {(mag)} & {(mag)}
}
\startdata
Mrk 709 S & 9:49:18.02 & +16:52:44.2 & 0.052 & $2.5 \times 10^9$ & $16.69\pm0.00$ & $16.77\pm0.00$ & $16.48\pm0.00$\\
Mrk 709 N & 9:49:18.10 & +16:52:49.5 & 0.052 & $1.1 \times 10^9$ & $17.12\pm0.01$ & $16.73\pm0.01$ & $16.38\pm0.01$ \\
\enddata
\tablecomments{The reported right ascension, declination, and $g$, $r,$ and $i$ photometric values are from the 16th Data Release (DR16) of the Sloan Digital Sky Survey \citep[SDSS-IV;][]{blanton2017,Ahumada2020}. The redshift of Mrk 709 N is measured in this work using Keck spectroscopy. The redshift of Mrk 709 S and the stellar masses are from \cite{reinesetal2014}.}
\end{deluxetable*}

\section{Observations} \label{sec:observations}

We obtained multi-wavelength imaging of Mrk 709 with the Wide Field Camera 3 (WFC3) aboard {\it HST}. The images were taken over 2 orbits on 2015 November 18-19 (Proposal ID 14047; PI Reines).  The galaxy was observed in three filters: a broad-band near-IR filter (IR/F110W), a medium-band optical filter (UVIS/F621M), and a narrow-band H$\alpha$ filter (UVIS/F680N). 
For each filter, we obtained four dithered exposures to facilitate cosmic-ray rejection, avoid bad pixels, and improve the PSF sampling. We obtained the calibrated and drizzled images produced by the STScI data reduction pipeline.  The near-IR image has a plate scale of 0\farcs13/pixel and the optical/H$\alpha$ images have plate scales of 0\farcs04/pixel. Our observations are summarized in Table 1.

\subsection{Hubble Space Telescope Imaging}

\begin{deluxetable}{cccc}
\tablecaption{HST observations\label{tab:observations}}
\tablewidth{0pt}
\tablehead{
\colhead{Filter} & \colhead{Instrument} & \colhead{Description} & \colhead{Exp. Time}
}
\startdata
F110W & WFC3/IR & Near IR; 1.15 $\mu$m & 400s\\
F621M & WFC3/UVIS & Optical; 0.62 $\mu$m & 1600s\\
F680N & WFC3/UVIS & redshifted H$\alpha$ & 2300s 
\enddata
\label{tab:obs}
\end{deluxetable}

The final three galaxy images were slightly misaligned because the Fine Guidance Sensors on {\it HST} could only acquire a single guide star during the observations, which allows a slow drift of the target on the detector.  Using the F621M image as the reference, we shifted the F110W image 0.60 pixels west and 0.36 pixels north, and the F680N image 0.84 pixels west and 0.19 pixels north.  We also corrected the absolute astrometry by comparing the adjusted {\it HST} images to SDSS images of Mrk 709 and updating the WCS. All three images were moved 0\farcs29 south and 0\farcs48 east to match the SDSS astrometry. We estimate the uncertainty in the astrometry of the {\it HST} images relative to the SDSS to be $\sim$0\farcs05. Figure 1 shows a 3-color {\it HST} image of Mrk 709.

\subsection{Keck Spectroscopy}

We observed Mrk 709 for 600s with the dual-beam Low Resolution Imaging Spectrometer \citep[LRIS;][]{Okeetal1995} on the Keck I 10-meter telescope on UT 2018 March 18. Our instrument setup consisted of the 1\arcsec\ width slit, the 5600 \AA\ dichroic, the 600 $\ell$ mm$^{-1}$ blue grism with $\lambda_{\rm blaze}=4000$ \AA, and the 400 $\ell$ mm$^{-1}$ red grating with $\lambda_{\rm blaze}=8500$ \AA.  
This configuration covers the full optical window at a resolving power of $R \approx 1100$.  We used standard data reduction techniques in IRAF and for flux calibration we used observations obtained on the same night with the same instrument configuration of the standard stars Hiltner~600 and Wolf~1346 from Massey \& Gronwall (1990; ApJ, 358, 344). The spatial resolution is 1\farcs2, due to the seeing on the night of the observation.

\begin{figure}
\includegraphics[width=3.5in]{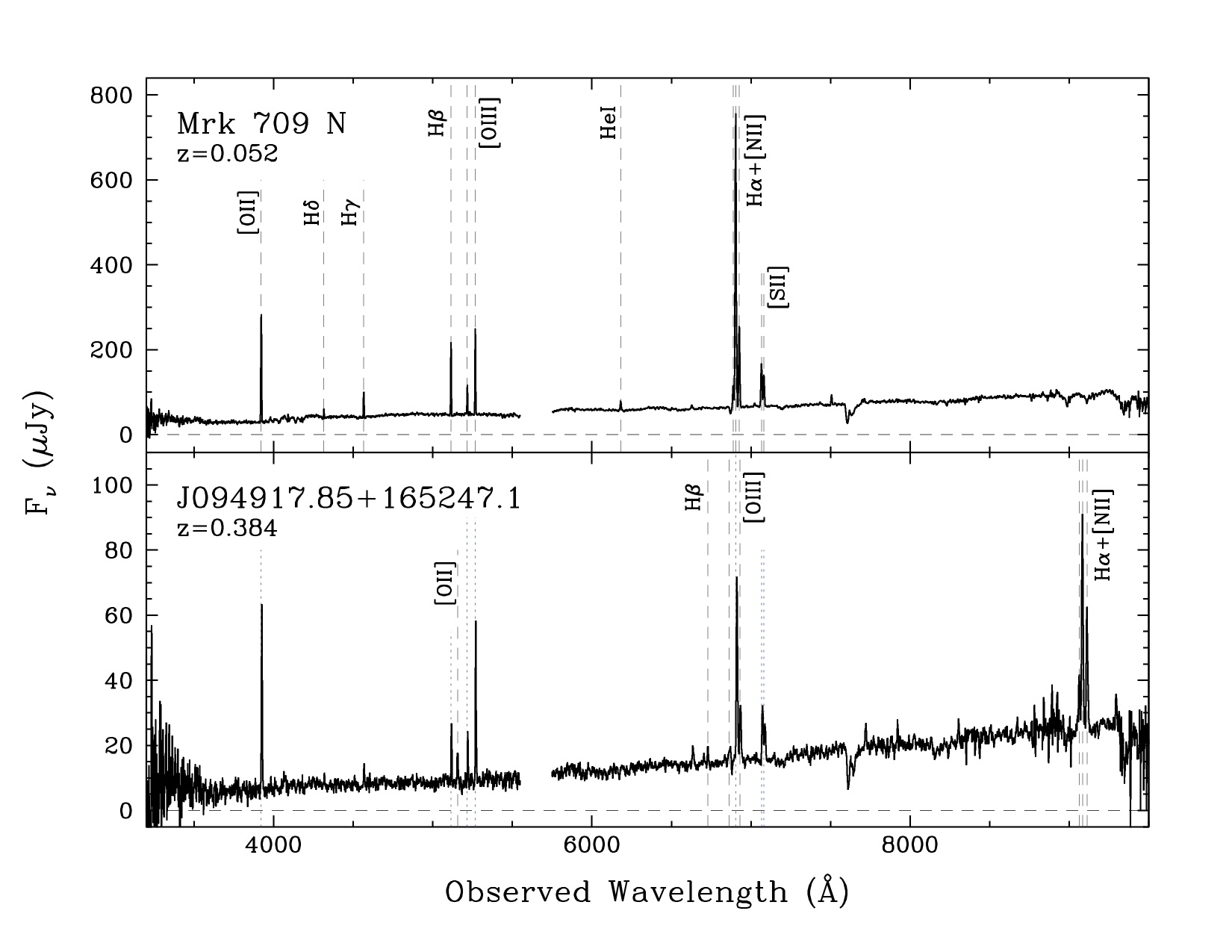}
\caption{Keck/LRIS spectra of Mrk 709 N (top panel) and J094917.85+165247.1, or the western barred spiral in Figure~\ref{fig: Fig 1} (bottom panel), with strong lines labeled. Mrk 709 N has the same redshift as Mrk 709 S ($z=0.052$; \citealt{reinesetal2014}). The western object is a background galaxy with a redshift $z=0.384$, and is therefore not included in the Mrk 709 system.}\label{fig:keckspec}
\end{figure}

We positioned the slit at a position angle of PA = 56.5\degr\ to cover both the northern and western galaxies, to measure their redshifts, and determine if they are part of the same system as Mrk 709 S.   The redshift of the southern galaxy, Mrk 709 S, is known to be $z=0.052$ from SDSS spectroscopy \citep{reinesetal2014}. Figure \ref{fig:keckspec} shows the reduced Keck/LRIS spectra using a 1\arcsec\ extraction width.  The spectra indicate that Mrk 709 N is at the same redshift as Mrk 709 S ($z=0.052$), but the western galaxy is at a higher redshift ($z=0.364$) and therefore not part of the Mrk 709 system.

\subsection{Very Long Baseline Interferometry}\label{sec:hsa}

We obtained VLBI observations of Mrk 709 S on February 14, 2015.  We used the HSA, comprising of the Very Long Baseline Array (VLBA) antennas along with the Green Bank Telescope (GBT) and the phased Very Large Array (VLA).  The project code is BR204. All VLBA antennas observed with the exception of Mauna Kea (MK) and St Croix (SC); Pie Town (PT) and Hancock (HN) were unavailable for a portion of the observation.  2 $\times$ 128 MHz subbands, centered at 1440 MHz and 1696 MHz, were sampled in dual polarisation.  The bright source 4C39.25 was observed twice to calibrate the instrumental bandpass, while JVAS~J0949+1752 was observed every 5 minutes as a phase reference.  The total observation duration was 5 hours. 

We made use of a pipeline implemented using the ParselTongue \citep{kettenis06a} python interface to AIPS \citep{greisen03a}, optimised for 1.5~GHz VLBI observations.  This pipeline has previously been used for numerous VLBA astrometric observations of radio pulsars, and a detailed description is available in \citet{deller19a}.  Briefly, it applies {\em a priori} amplitude calibration, followed by bandpass correction derived from 4C39.25, and then fringe-fitting and amplitude self-calibration using JVAS~J0949+1752.  

After calibration, the data were imaged using difmap \citep{shepherd97a}.  Natural weighting was used, resulting in a synthesised beam size of $7 \times 23$ mas.  We produced a $1.5'' \times 1.5''$ image covering the entire region of Mrk 709 S, with an image rms ($\sigma$) of 4 $\mu$Jy.  No source was visible in the dirty image with a peak brightness above 4.5$\sigma$, and we set a 5$\sigma$ upper limit to the peak brightness of any 1.5 GHz radio emission to be $\leq20\mu$Jy/beam.


\begin{deluxetable*}{ccccccc}
\tablecaption{{Star Formation Rates}\label{tab:sfr}}
\tablehead{
\colhead{Component} & \colhead{RA} & \colhead{Dec} & \colhead{Aperture Size} & \colhead{PA} & \colhead{H$\alpha$ Flux} & \colhead{SFR}\\
{} & & & {(arcsec $\times$ arcsec)} & {(deg)} & {($10^{-14}$~erg~s$^{-1}$~cm$^{-2}$)} & {($M_\odot$~yr$^{-1}$)}}
\startdata
Mrk 709 S & 9:49:18.03 & +16:52:44.2 & $1.5 \times 1.0$ & 110 & $3.15\pm0.03$ & 1.0 \\
Mrk 709 N & 9:49:18.11 & +16:52:49.5 & $1.5 \times 1.5$ & 0 & $0.46\pm0.02$ & 0.2 \\
Bridge & 9:49:18.06 &  +16:52:46.7 & $6.5 \times 0.75$ & 52 & $1.6\pm0.04$ & 0.5\\
\enddata
\tablecomments{Star formation rates of the different components of the Mrk 709 system are derived from our {\it HST} imaging. We use elliptical apertures with the given semi-major and semi-minor axes, and position angles (PA).  
These SFRs should be considered lower limits (see Section \ref{sec:system}).}
\label{tab:components}
\end{deluxetable*}

\section{The Mrk 709 System}\label{sec:system}

Figure 1 shows the {\it HST} images of Mrk 709.  Three galaxies are apparent in the field of view.
The northern and southern galaxies, Mrk 709 N and Mrk 709 S, were identified by \citet{reinesetal2014} using SDSS imaging, however only Mrk 709 S had an SDSS spectrum, which indicated a redshift of $z=0.052$ ($d \sim 214$ Mpc).  Our new long-slit Keck spectroscopy covers the northern and western galaxies and indicates redshifts of $z=0.052$ and $z=0.384$, respectively. Therefore, while Mrk 709 N and Mrk 709 S are part of the same system, the western component is a background galaxy (J094917.85+165247.1) and not part of the Mrk 709 system. We present some of the basic galaxy properties for both Mrk 709 N and Mrk 709 S in Table~\ref{tab:properties}.

There is a striking bridge of star clusters between Mrk 709 N and Mrk 709 S, strongly emitting in H$\alpha$. The bridge is a characteristic tidal feature indicating that the two dwarf galaxies are undergoing a merger/interaction. The bridge is $\sim 10$\arcsec\ long. At the distance of Mrk 709, this corresponds to a projected physical length of $\sim 10$ kpc. 


\begin{figure*}[!t]
\begin{center}
\includegraphics[width=0.8\textwidth]{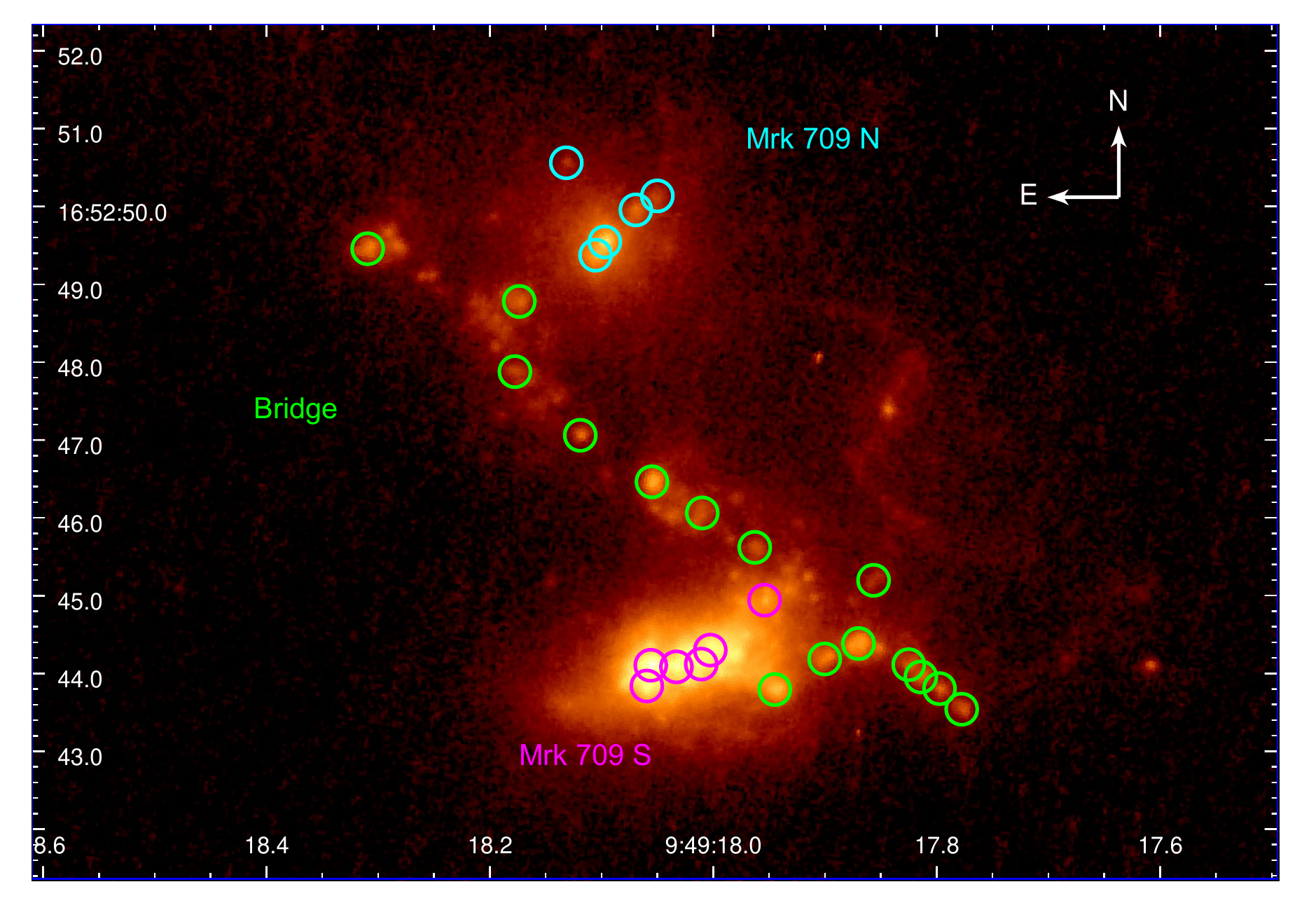}
\caption{{\it HST}/F680N image of Mrk 709 showing the observed H$\alpha$ emission and continuum. Mrk 709 N, Mrk 709 S and the bridge associated with Mrk 709 system are labeled. We over-plot the 0\farcs2 apertures used to perform photometry on the identified star clusters, where those in Mrk 709 N are shown in cyan, those in Mrk 709 S are shown  in magenta and those in the bridge are shown in green.}\label{fig:locations}
\end{center}
\end{figure*}

We estimate the star formation rates (SFRs) of Mrk 709 S, Mrk 709 N and the bridge using H$\alpha$ fluxes determined from our {\it HST} images and the following relation from \cite{Kennicutt2012}:

\begin{equation}
   {\rm log~SFR}~(M_\odot~{\rm yr}^{-1}) = {\rm log}~L_{\rm H\alpha} - 41.27,
\end{equation} 

\noindent
where $L_{\rm H\alpha}$ is the luminosity of H$\alpha$ in units of erg s$^{-1}$. The H$\alpha$ flux is given by:

\begin{equation}
    F_{\rm H\alpha} = (f_{\rm F680N} - f_{\rm cont})\Delta\lambda_{\rm F680N}, 
\label{eqn:haflux}
\end{equation}

\noindent
where $f_{\rm F680N}$ is the flux density in the F680N image, $f_{\rm cont}$ is the continuum flux density at the wavelength of redshifted H$\alpha$, and $\Delta\lambda_{\rm F680N}$ is the width of the F680N filter.  The continuum is found by interpolating between the F621M and F110W flux densities in log $f_\lambda$ vs.\ log $\lambda$ space since a power law is a good approximation of a stellar population spectrum redward of $\sim 4000$\AA\ \citep{Leitherer1999}. 
Table \ref{tab:sfr} presents the measurements and aperture definitions.

We find SFRs of approximately 1, 0.2, and 0.5 $M_\odot$ yr$^{-1}$ for Mrk 709 S, Mrk 709 N, and the bridge, respectively.  These values should be taken as lower limits since the measured H$\alpha$ fluxes cannot be corrected for extinction with the data in hand. It is also possible that our continuum estimates are artificially high due to nebular emission contaminating the broadband filters \citep{reinesetal2010}. While the redshifted [NII]$\lambda \lambda6548,6584$ lines fall in the F680N filter along with H$\alpha$, the contribution to the total flux and SFR is expected to be minor ($\sim 15\%$ based on the SDSS spectrum of Mrk 709 S). We do not apply a correction to the measured SFRs as we do not know the contamination from the [NII]$\lambda \lambda6548,6584$ lines for every galaxy component.

Our SFR for Mrk 709 S is somewhat less than the value of 3.9 $M_\odot$ yr$^{-1}$ found by \citet{reinesetal2014} using the SDSS spectrum, which was taken in a 3\arcsec\ aperture centered on Mrk 709 S and corrected for extinction using the Balmer decrement (H$\alpha$/H$\beta$=3.47; \citealt{reinesetal2014}). 
The value derived here is likely lower due to the combination of a smaller aperture, no extinction correction, and uncertainty in the underlying stellar continuum using photometric measurements.
We also note that both \citet{reinesetal2014} and this work assume that the AGN in Mrk 709 S does not contribute significantly to the observed H$\alpha$ emission since the line ratios suggest that star formation dominates. 
The derived SFRs are very high given the low stellar masses of only $M_\star \sim 2.5 \times 10^9 M_\odot$ and $M_\star \sim 1.1 \times 10^9 M_\odot$ for Mrk 709 S and Mrk 709 N, respectively \citep{reinesetal2014}. Mrk 709~S has an SFR at least as large as the Milky Way.

\section{Star Forming Regions}\label{sec:sfreg}

With the dramatically improved spatial resolution afforded by {\it HST} (e.g., compared to the SDSS images presented in \citealt{reinesetal2014}), we are now in a position to investigate individual star forming regions and young star clusters within the Mrk 709 system.  The H$\alpha$ and continuum images, along with population synthesis models, enable us to estimate the ages, ionizing luminosities, and stellar masses for each star forming region/cluster. 

\subsection{Photometry of Star Clusters}

\begin{deluxetable*}{cccccc}
\tablecaption{Photometry of Star clusters in Mrk709\label{tab:photometry}}
\tablewidth{0pt}
\tablehead{
\colhead{Source} & \colhead{RA} & \colhead{Dec} & \colhead{F680N} & \colhead{F621M} & \colhead{F110W}
}
\startdata
\cutinhead{Mrk 709 S}
1s & 9:49:17.94 & +16:52:43.8 & 21.41(0.16) & 21.50(0.17) & 22.34(0.31)\\
2s & 9:49:18.00 & +16:52:44.3 & 19.61(0.12) & 19.63(0.06) & 20.67(0.13)\\
3s & 9:49:18.01 & +16:52:44.1 & 19.63(0.18) & 19.78(0.14) & 20.71(0.19)\\
4s & 9:49:18.03 & +16:52:44.1 & 19.54(0.13) & 19.70(0.10) & 20.51(0.23)\\
5s & 9:49:18.06 & +16:52:44.1 & 19.07(0.09) & 19.19(0.04) & 20.13(0.06)\\
6s & 9:49:18.06 & +16:52:43.8 & 19.60(0.08) & 20.32(0.07) & 21.08(0.12)\\
\cutinhead{Mrk 709 N}
1n & 9:49:18.05 & +16:52:50.1 & 23.11(0.29) & 23.69(0.50) & 24.39(0.92)\\
2n & 9:49:18.07 & +16:52:50.0 & 22.63(0.55) & 23.13(0.59) & 23.84(0.76)\\
3n & 9:49:18.10 & +16:52:49.5 & 20.71(0.17) & 21.36(0.20) & 22.14(0.36)\\
4n & 9:49:18.11 & +16:52:49.4 & 20.96(0.17) & 21.69(0.23) & 22.38(0.42)\\
5n & 9:49:18.13 & +16:52:50.6 &  23.83(0.10) & 24.56(0.23) & 26.26(3.62)\\
\cutinhead{Bridge}
1b & 9:49:17.78 & +16:52:43.5 & 22.86(0.03) & 23.81(0.05) & 25.24(0.04)\\
2b & 9:49:17.80 & +16:52:43.8 & 22.77(0.07) & 23.83(0.04) & 24.55(0.02)\\
3b & 9:49:17.81 & +16:52:44.0 & 22.41(0.07) & 24.35(0.19) & 24.79(0.03)\\
4b & 9:49:17.83 & +16:52:44.1 & 22.49(0.08) & 23.99(0.04) & 24.95(0.24)\\
5b & 9:49:17.86 & +16:52:45.2 & 23.52(0.05) & 24.70(0.07) & 25.83(0.64)\\
6b & 9:49:17.87 & +16:52:44.4 & 21.52(0.06) & 23.09(0.04) & 24.24(0.14)\\
7b & 9:49:17.90 & +16:52:44.2 & 22.26(0.06) & 23.28(0.11) & 26.30(2.43)\\
8b & 9:49:17.94 & +16:52:43.8 & 21.00(0.01) & 22.28(0.04) & 23.41(0.36)\\
9b & 9:49:17.96 & +16:52:45.6 & 23.25(0.07) & 24.70(0.05) & 25.28(0.23)\\
10b & 9:49:18.01 & +16:52:46.1 & 22.77(0.03) & 24.58(0.06) & 25.40(0.32)\\
11b & 9:49:18.05 & +16:52:46.5 & 21.60(0.02) & 22.95(0.03) & 24.03(0.04)\\
12b & 9:49:18.12 & +16:52:47.1 & 23.01(0.03) & 25.72(0.38) & 26.62(0.18)\\
13b & 9:49:18.17 & +16:52:48.8 & 22.72(0.12) & 23.55(0.22) & 24.47(0.60)\\
14b & 9:49:18.18 & +16:52:47.9 & 23.33(0.45) & 25.43(0.77) & 25.40(0.43)\\
15b & 9:49:18.31 & +16:52:49.5 & 22.07(0.02) & 23.46(0.04) & 24.73(0.03)
\enddata
\tablecomments{Magnitudes are given in the STMAG photometric system. 
}
\label{tab:clusterphot}
\end{deluxetable*}

We identify young star clusters in the narrowband H$\alpha$ (F680N) image. The regions are shown in Figure \ref{fig:locations} and their coordinates are listed in Table \ref{tab:clusterphot}. Using the astropy and photutils packages in Python, we perform aperture photometry on a total of 26 clusters; six in Mrk 709 S, five in Mrk 709 N, and 15 in the bridge between the two dwarf galaxies. We used circular apertures of radius 0\farcs2. The background was estimated within an annulus of inner radius 0\farcs45 and outer radius 0\farcs65 centered on each cluster. However, we made multiple measurements varying the background since this is the dominant source of uncertainty in our photometry.  Aperture corrections were applied to our final background-subtracted flux densities using the encircled energy fractions for a 0\farcs2 aperture: 0.839, 0.847, and 0.712 for the F680N, F621M, and F110W images respectively. 
The final magnitudes are listed in Table \ref{tab:photometry} and given in the STMAG photometric system\footnote{https://hst-docs.stsci.edu/wfc3dhb/chapter-9-wfc3-data-analysis/9-1-wfc3-data-analysis}.

\subsection{Physical Properties of the Clusters}
\label{ssec:phys_prop}
Estimates of the physical properties of the star clusters come from comparing our measurements to Starburst99 (v5.1) population synthesis models \citep{Leitherer1999}. We use a simulation of an instantaneous burst of $10^5~M_\odot$ with a Kroupa initial mass function.  We adopt the Geneva evolutionary tracks with high mass loss, a metallicity of $Z=0.001$ and the Pauldrach/Hillier atmospheres. The model metallicity is similar to that of Mrk 709 \citep[$\sim10$\% solar;][]{masegosa1994}.


\subsubsection{Ages}

We estimate the ages of star clusters by comparing the equivalent width of H$\alpha$ emission to the model predictions as a function of age.  The equivalent width of H$\alpha$ is given by the ratio of the flux of the H$\alpha$ emission line (which is proportional to the ionizing flux from short-lived massive stars) to the continuum flux density at the same wavelength:

\begin{equation}
    W_{\rm H\alpha} = \frac{F_{\rm H\alpha}}{f_{\rm cont}}.
    \label{ew_ha}
\end{equation}  

\noindent
This ratio is strongly dependent on age for stellar clusters between $\sim$ 3 and 20 Myr old. The insensitivity before $\lesssim$3 Myr occurs because the most massive stars have not yet died, thereby reducing the ionizing flux, and the insensitivity after $\sim 20$ Myr occurs because clusters older than this are not strong H$\alpha$ emitters.  The H$\alpha$ flux is given by Equation \ref{eqn:haflux} and determined for individual clusters using the same procedure as described in Section \ref{sec:system}.

\begin{figure}
    \centering
    \plotone{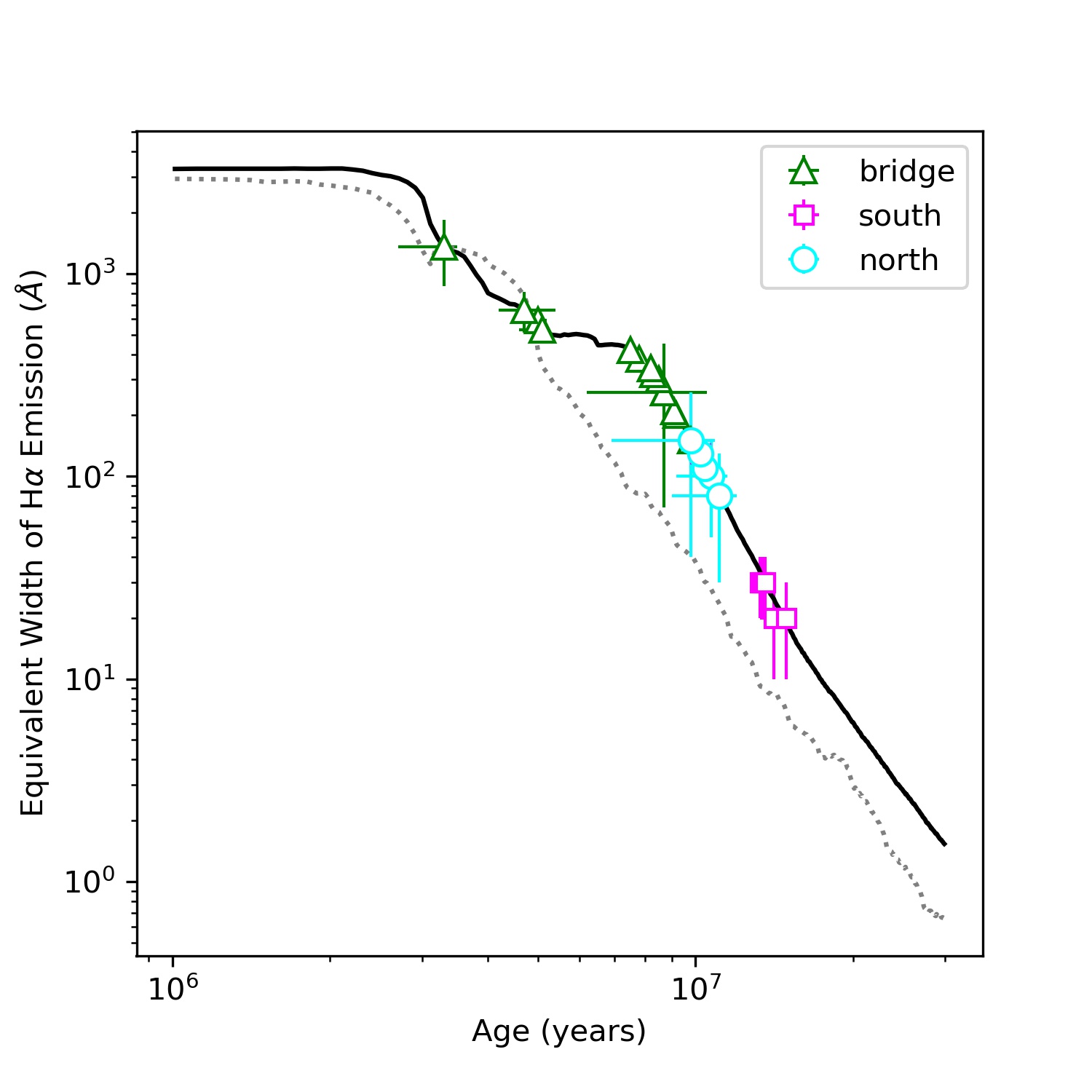}
    \caption{Starburst99 model evolutionary tracks for H$\alpha$ equivalent width. The solid black line shows our fiducial model with $Z=0.001$.  We also show a model with $Z=0.004$ (dotted line) to illustrate the age-metallicity degeneracy. The star clusters in the bridge are denoted by green triangles, those in Mrk 709 S are denoted by magenta squares and those in Mrk 709 N are denoted by cyan circles. The errors in age 
    represent the ages associated with the minimum and maximum 
    H$\alpha$ equivalent width using our fiducial model. The clusters in each galaxy and the bridge are grouped together, and the bridge hosts the youngest star clusters.}
    \label{fig:ha_ew}
\end{figure}

In Figure \ref{fig:ha_ew} we show the measured H$\alpha$ equivalent width of star clusters in the Mrk 709 system as well as the model predictions as a function of age.  
The southern galaxy has the oldest star clusters, with an average age of 13.3 Myr. Star clusters in the northern galaxy have an average age of 10.4 Myr, and the star clusters in the bridge have an average age of 7.5 Myr. The star clusters in the bridge have the largest equivalent widths and are the youngest in the system. 

We consider the derived ages to be upper limits for a couple of reasons.  First, as described in Section \ref{sec:system}, the continuum may be overestimated causing the H$\alpha$ equivalent width to be underestimated.  Second, the model metallicity is a factor of two lower than the measured value for Mrk 709. Given the known age-metallicity degeneracy, this may lead to an overestimate of the ages of the star clusters.  To illustrate this, we show an additional model with $Z=0.004$ in Figure \ref{fig:ha_ew}.

\begin{deluxetable*}{cccccc}
\tablecaption{Physical Properties of Star Clusters in Mrk 709
\label{tab:physprop}}
\tablewidth{0pt}
\tablehead{
\colhead{Source} &  \colhead{H$\alpha$ flux} & \colhead{H$\alpha$ EW} & \colhead{Age} & \colhead{log $M_\star$} & \colhead{Q$_{\rm Lyc}$} \\ \colhead{} & \colhead{(10$^{-16}$ erg s$^{-1}$ cm$^{-2}$)} & \colhead{({\AA})} & \colhead{(Myr)} & \colhead{($M_\odot$) } & \colhead{(10$^{49}$ s$^{-1}$)}
}
\startdata
\cutinhead{Mrk 709 S}
1s & 2.1(0.3) & 20(10) & (0.5)14.1(0.8) & 6.9(0.3) & 90(10)\\
2s & 9.0(1.2) & 20(10) & (0.5)14.9(0.5) & 7.7(0.2) & 390(50)\\
3s & 13.9(1.8) & 30(10) & (0.5)13.3(0.5) & 7.6(0.3) & 600(80)\\
4s & 14.3(1.4) & 30(10) & (0.3)13.4(0.4) & 7.6(0.2)& 620(60)\\
5s & 21.1(1.4) & 30(10) & (0.2)13.6(0.2) & 7.9(0.1) & 910(60)\\
6s & 31.9(0.8) & 130(10) & (0.1)10.2(0.1) & 7.3(0.1) & 1380(40) \\
\cutinhead{Mrk 709 N}
1n & 1.1(0.1) & 100(50) & (0.8)10.7(1.5) & 6.0(0.6) & 50(10)\\
2n & 1.5(0.3) & 80(50) & (0.9)11.1(2.1) & 6.2(0.7) & 70(10)\\
3n & 10.9(0.6) & 110(20) & (0.3)10.4(0.4) & 6.9(0.2) & 470(30)\\
4n & 9.1(0.5) & 130(20) & (0.4)10.2(0.4) & 6.8(0.2) & 390(20)\\
5n & 0.7(0.1) & 150(110) & (1.1)9.8(2.9) & 5.6(0.9) & 30(10) \\
\cutinhead{Bridge}
1b & 1.9(0.1) & 200(20) & (0.1)9.2(0.2) & 5.8(0.1) & 80(10)\\
2b & 2.1(0.1) & 210(10) & (0.0)9.1(0.1) & 5.8(0.1) & 90(10)\\
3b & 3.7(0.1) & 590(50) & (0.2)5.0(0.1) & 5.1(0.2) & 160(10)\\ 
4b & 3.2(0.1) & 380(20) & (0.1)7.8(0.2) & 5.7(0.1) & 140(10)\\
5b & 1.1(0.1) & 260(20) & (0.1)8.7(0.2) & 5.5(0.1) & 50(10)\\
6b & 8.0(0.1) & 420(20) & (0.2)7.5(0.2) & 6.0(0.1) & 350(10)\\
7b & 3.8(0.1) & 270(130) & (1.0)8.6(1.3) & 6.0(1.3) & 160(10)\\
8b & 12.0(0.1) & 300(10) & (0.1)8.5(0.1) & 6.4(0.1) & 520(10)\\
9b & 1.5(0.1) & 340(10) & (0.1)8.2(0.1) & 5.5(0.1) & 70(10)\\
10b & 2.6(0.1) & 530(30) & (0.0)5.1(0.5) & 5.0(0.1) & 110(10)\\
11b & 7.1(0.1) & 320(10) & (0.1)8.3(0.1) & 6.1(0.1) & 310(10)\\
12b & 2.4(0.1) & 1360(490) & (0.2)3.3(0.6) & 4.5(0.3) & 100(10)\\ 
13b & 2.0(0.1) & 150(40) & (0.5)9.8(0.5) & 6.0(0.3) & 80(10)\\
14b & 1.6(0.1) & 660(150) & (0.7)4.7(0.5) & 4.6(0.5) & 70(10)\\ 
15b & 4.7(0.1) & 340(20) & (0.3)8.2(0.1) & 6.9(0.1) & 200(10)
\enddata
\tablecomments{All uncertainties in this table are shown in parentheses. The H$\alpha$ fluxes and EWs are measured from the {\it HST} imaging and are not corrected for extinction. The errors on the stellar masses do not include a systematic uncertainty of $\sim 0.3$ dex due to uncertainties in stellar evolution \citep{conroyetal2009}.}
\label{tab:prop}
\end{deluxetable*}

\subsubsection{Ionizing Luminosities}

We estimate the production rate of ionizing photons Q$_{Lyc}$ from the star-forming regions using the following equation derived from \cite{condon1992} for a $10^4$ K gas: 

\begin{equation}
    \left(\frac{Q_{\rm Lyc}}{\rm s^{-1}} \right) \gtrsim 7.87 \times 10^{11} \left( \frac{ L_{\rm H\alpha}}{\rm erg~s^{-1}} \right).
\label{eqn:Q}
\end{equation}

\noindent
This only provides a lower limit on the ionizing luminosity since some photons may be absorbed by dust or escape the region before ionizing hydrogen atoms.   

Adopting a distance of 214 Mpc, we find that the ionizing luminosities are in the range of $3 \times 10^{50}$ to $1.4 \times 10^{52}~\rm s^{-1}$. Assuming a ``typical" O-type star (type O7.5 V) produces $Q_{\rm Lyc} = 10^{49}$ s$^{-1}$ \citep{Vacca1996}, $\sim$ 30 to 1400 equivalent O stars are necessary to power the ionizing luminosities of the star clusters in Mrk 709, which is comparable to young massive star clusters in other dwarf starburst galaxies \citep[e.g.,][]{reinesetal2008a,reinesetal2008b}. 

\subsubsection{Stellar Masses}\label{sec:masses}

We estimate the stellar masses of young star clusters using the ionizing luminosities and ages derived above. 
The models predict the number of Lyman continuum photons as a function of age given an input total stellar mass ($M_\star = 10^5~M_\odot$ in the model used here).  Assuming stellar mass scales linearly with the Lyman continuum photon production, we use the model prediction of $Q_{\rm Lyc}$ at the age of a cluster and multiply the input model mass by the ratio of the observed to the model ionizing luminosity. The stellar masses estimated using this method are in the range of $M_\star \sim 10^5-10^8~M_\odot$ (see Table \ref{tab:prop}).

The star clusters in Mrk 709 S are extremely massive with an average mass of $M_\star \sim 4 \times 10^7 M_\odot$.  This is significantly larger than young massive star clusters in the local Universe (typically $\lesssim 10^6 M_\odot$; \citealt{Whitmore2010}, \citealt{reinesetal2008b}) and the star formation in Mrk 709 S is more akin to that seen in higher redshift irregular galaxies (i.e., clump cluster and chain galaxies, \citealt{Elmegreen2009}).  Both the clusters in Mrk 709 S and the clumps in these higher redshift galaxies have typical masses of $10^7-10^8 M_\odot$ and each contain an average of $\sim 2\%$ their galaxy mass.  The total mass in clumps relative to total galaxy mass is also similar at $\sim 10\%$.  The size scales are slightly different, however, with star forming regions in clump clusters and chain galaxies having typical sizes of $\sim 1.8$ kpc \citep{Elmegreen2005}.  This is comparable to the size of brightest central region in Mrk 709 S containing $\sim 5$ massive clusters.




The star clusters in Mrk 709 N and the bridge are also massive, but not as extreme as in Mrk 709 S.  The average masses are $\sim 3 \times 10^6 M_\odot$  and $\sim 7 \times 10^5 M_\odot$, for Mrk 709 N and the bridge respectively, which are comparable to massive globular clusters.  The total cluster mass in Mrk 709 N is $M_\star \sim 1.7 \times 10^7~M_\odot$, roughly 2\% of the total galaxy mass ($M_\star \sim 1.1 \times 10^9~M_\odot$; \citealt{reinesetal2014}).  

\begin{figure}
    \hspace{-.2in}     
    \includegraphics[width=3.95in]{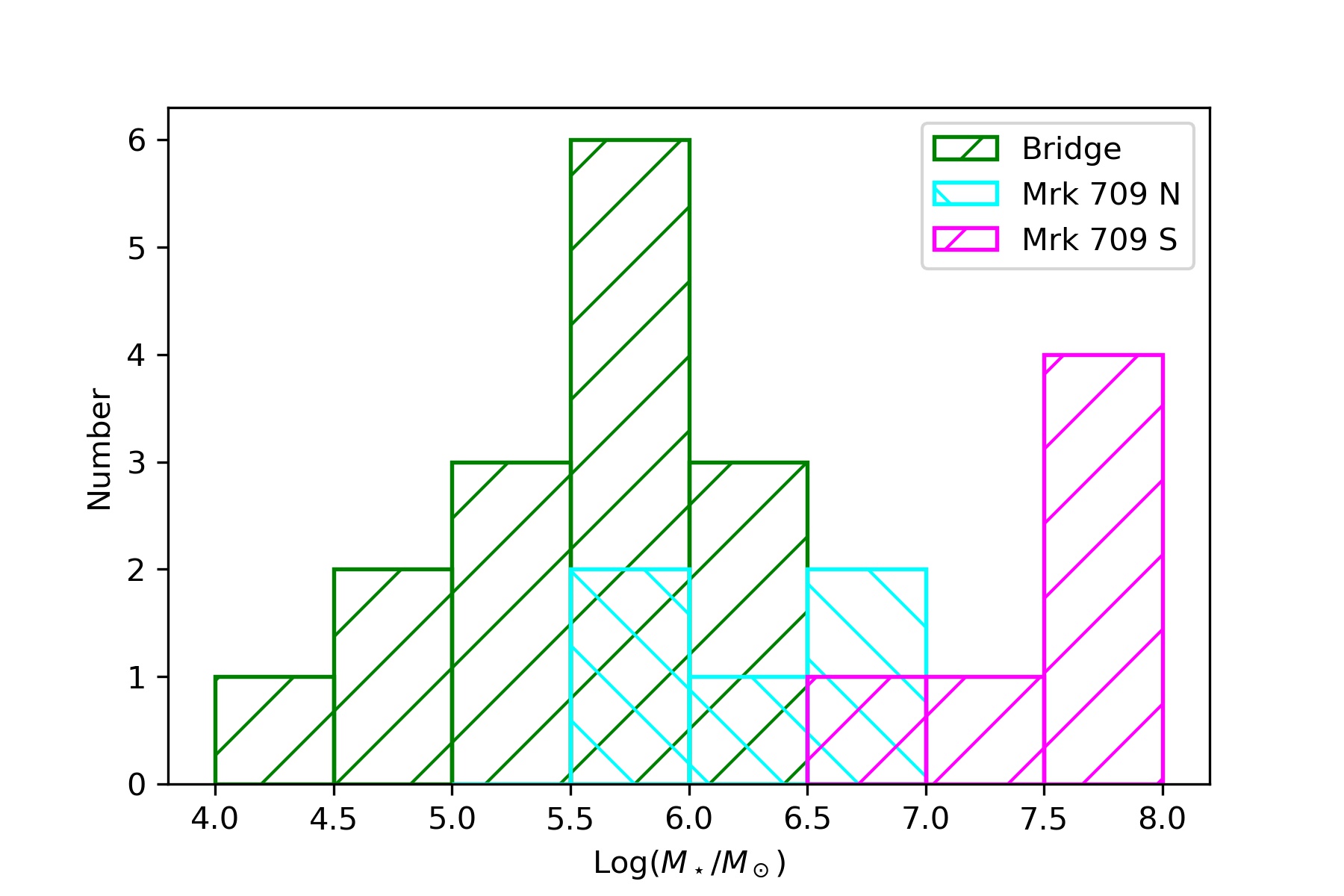}
    \caption{The mass distribution of star clusters in Mrk 709 N (cyan), Mrk 709 S (magenta) and the bridge (green). The histogram shows the large spread of masses in the star clusters. Bridge star clusters tend to lower masses $ 10^{4-6} M_\odot$ and star clusters in Mrk 709 S tend to higher masses $ 10^{6.5-8} M_\odot$. }
    \label{fig:mass_hist}
\end{figure}

\section{The Massive Black Hole in Mrk 709 S}\label{sec:agn}

\citet{reinesetal2014} identified a candidate massive BH in Mrk 709 S based on high-resolution radio and X-ray observations from the VLA and {\it Chandra}, respectively.  They find a compact ($\lesssim 0\farcs3$) radio source with a 5 GHz luminosity of $\nu L_\nu = (1.6 \pm 0.6) \times 10^{37}$ erg s$^{-1}$ that is consistent with the position of a hard X-ray point source with $L_{\rm 2-10 keV} \ge (5.0 \pm 2.9) \times 10^{40}$ erg s$^{-1}$. The offset between the radio and X-ray sources is $0\farcs17$, which is within the astrometric uncertainties. Assuming the radio and X-ray emission are indeed coming from the same source, \citet{reinesetal2014} argue for a BH origin and estimate a mass in the range of $M_{\rm BH} \sim 10^{5-7}~M_\odot$ using the fundamental plane of BH activity \citep[e.g.,][]{merlonietal2003, falckeetal2004, plotkinetal2012, gultekinetal2019}. A stellar-mass X-ray binary is firmly ruled out by the high radio luminosity.  

\subsection{A Radio-Detected Low-Luminosity AGN}\label{ssec:xrad}

\begin{figure}
    \centering
    \includegraphics[width=3.4in]{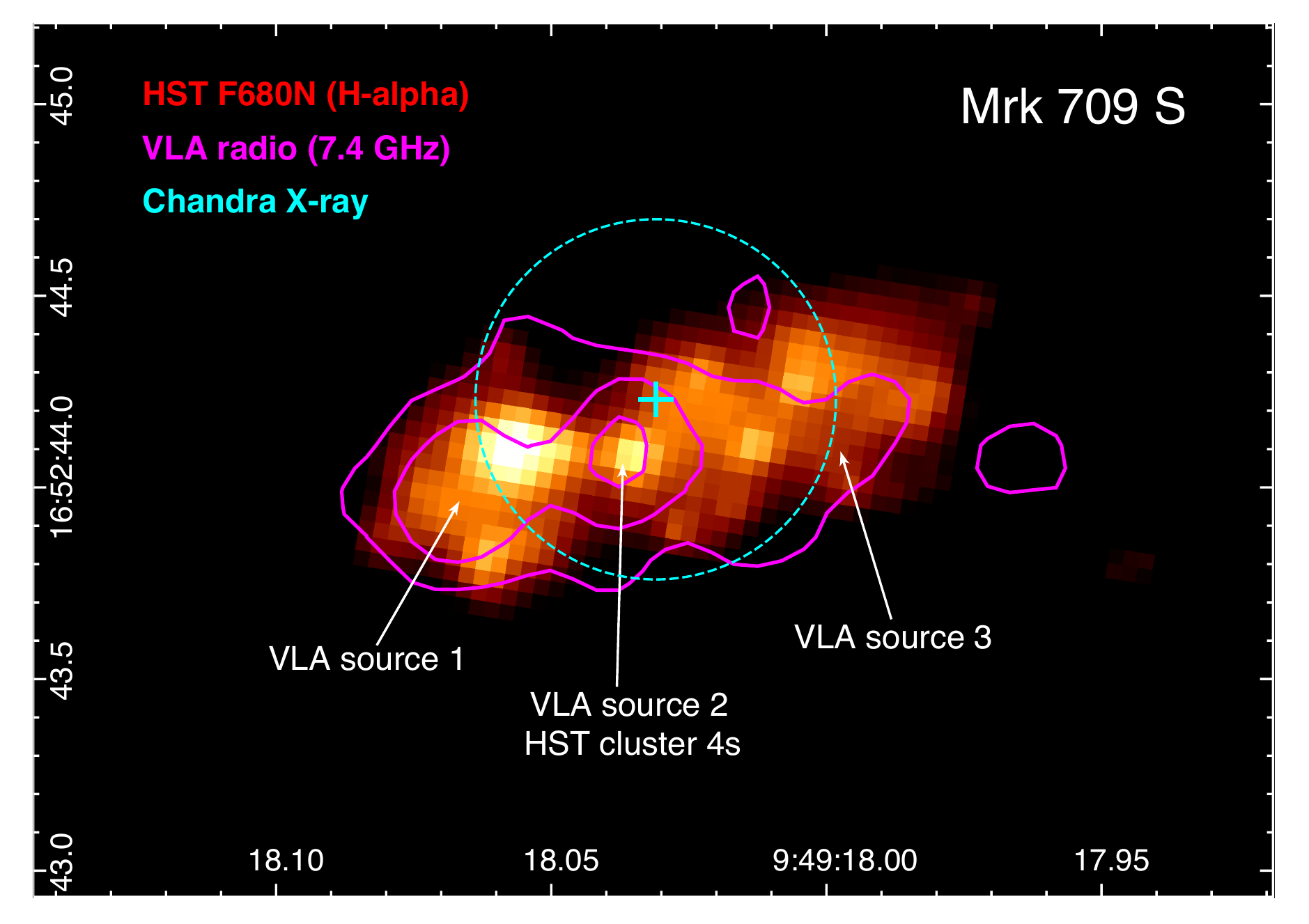}
    \caption{Close-up multi-wavelength view of Mrk 709 S.  The {\it HST} F680N image is shown in color with the stretch adjusted so that individual star clusters are visible.  We also show radio contours (magenta) and the position of the X-ray point source (and positional uncertainty; cyan) from the VLA and {\it Chandra} observations presented in \citet{reinesetal2014}.  \citet{reinesetal2014} associated a candidate massive BH with VLA radio source 2 as well as the X-ray source. The new {\it HST} observations presented here indicate that star cluster 4s is spatially coincident with VLA source 2, and consistent with the position of the X-ray source.}
    \label{fig:multiwave}
\end{figure}

Given that we cannot definitively say whether or not the radio and X-ray emission are indeed coming from the same physical source, we also consider separate origins for the radio and X-ray sources guided by our new {\it HST} imaging.  Figure \ref{fig:multiwave} shows that the X-ray source has a position consistent with multiple star clusters in Mrk 709 S, and could be possibly be an exceptionally luminous high-mass X-ray binary \citep{lehmeretal2019} provided the radio emission is not associated with the X-ray source.  From Figure \ref{fig:multiwave}, we also see that radio source 2 is spatially coincident with star cluster 4s (see Tables \ref{tab:clusterphot} and \ref{tab:prop}).  We therefore investigate whether the radio emission could plausibly come from a thermal H{\footnotesize{II}} region or have a supernova (SNe) origin following \citet{reinesetal2020}. 

Under the assumption that radio source 2 is an H{\footnotesize{II}} region associated with star cluster 4s, we can calculate the expected H$\alpha$ flux and compare this value to the measured value of $F_{\rm H\alpha} = 1.43 \times 10^{-15}$ erg s$^{-1}$ cm$^{-2}$ (Table \ref{tab:prop}).  Taking the 7.4 GHz flux density of 40 $\mu$Jy given by \citet{reinesetal2014} and using Equations 3 and 4a in \citet{condon1992} with an electron temperature of $T_e=10^4$~K, we expect an H$\alpha$ flux of $F_{\rm H\alpha} = 3.91 \times 10^{-14}$ erg s$^{-1}$ cm$^{-2}$ (modulo extinction) if the radio emission is thermal.  This is a factor of $27\times$ higher than the measured H$\alpha$ flux, which is exceptionally large even accounting for extinction due to dust. \citet{reinesetal2008a} and \citet{reinesetal2008b} find that this factor is $\lesssim 3$ in their studies of thermal radio emission associated with young, optically-visible star clusters in dwarf starburst galaxies.  We therefore conclude that radio source 2 is dominated by non-thermal emission. This is consistent with the radio spectrum of source 2 \citep{reinesetal2014}, which shows flux density decreasing with frequency.  

Next we show that the non-thermal radio emission from source 2 is unlikely to come from supernova remnants (SNRs) or younger SNe.  Using the relation given by \citet{chomiukwilcots2009} between the luminosity density of the brightest SNR/SNe in a galaxy and the SFR of the galaxy (and adopting the larger SFR of $3.9~M_\odot$ yr$^{-1}$ from \citealt{reinesetal2014}), we expect the most luminous SNR/SNe in Mrk 709 S to have $L_{\rm 1.4GHz}^{\rm max} \sim 3.6 \times 10^{19}$ W Hz$^{-1}$.  The spectral luminosity at 1.4 GHz of radio source 2 is estimated by \citet{reinesetal2014} to be $L_{\rm 1.4GHz} \sim 7 \times 10^{20}$ W Hz$^{-1}$ assuming a spectral index of $\alpha=-0.7$ ($S_\nu \propto \nu^\alpha$).\footnote{While the spectral index is not well constrained and could be in the range of steep to flat, it is not inverted \citep{reinesetal2014}.}  Given that this value is approximately $\sim 19\times$ higher than the expected luminosity of the brightest SNR/SNe, it is highly unlikely that even multiple SNRs/SNe in cluster 4s could account for the radio emission. We therefore conclude that the most plausible explanation for the origin of radio source 2 is a massive BH.  

This interpretation would have been further supported by the discovery of milliarcsecond-scale radio emission from source 2, which would have provided direct evidence of high brightness-temperature emission.  However, the non-detection in our VLBI HSA observations (see \S\ref{sec:hsa}) is still compatible with a massive BH.  The angular resolution of the HSA observations is almost two orders of magnitude finer than that of the VLA observations, but the sensitivities are only comparable (4 $\mu$Jy/beam vs.\ 5.2 $\mu$Jy/beam).  If source 2 is partially resolved into a core-jet object on milliarcsecond scales (which, given the VLA radio morphology, appears likely) the peak brightness could fall well below our HSA point source detection limit of $\sim 20~\mu$Jy ($5\sigma$) at 1.4 GHz.  The partial resolution of AGN emission in this way is commonplace: \citet{deller14a} performed a VLBI survey of $\sim$25,000 mJy radio sources (which should be dominated by radio AGN) detected in the VLA FIRST survey \citep{Becker1995} and found that only 30--40\% exhibited a milliarcsecond-scale core with a peak brightness $\geq32$\% of the arcsecond-scale flux density. 

Figure \ref{fig:multiwave} shows that VLA radio source 2 is surrounded by additional radio emission in Mrk 709 S.  The radio morphology is suggestive of lobes/outflow from the central AGN and the radio luminosity supports this scenario. We find that the measured radio luminosity is more than an order of magnitude larger than that expected from the combination of both thermal emission from H{\footnotesize{II}} regions and a population of SNRs/SNe in Mrk 709 S. Based on the H$\alpha$ flux ($F_{\rm H\alpha} = 3.15 \times 10^{-14}$ erg s$^{-1}$ cm$^{-2}$) of Mrk 709 S measured in the same elliptical aperture (1\farcs5 $\times$ 1\farcs0, PA = 110$\degr$) as the radio emission, we expect the contribution from thermal radio emission to be $L_{\rm 5 GHz} = 1.8 \times 10^{20}$ W Hz$^{-1}$ \citep{condon1992}.  Following \citet{reinesetal2020} and using the results of \citet{chomiukwilcots2009}, we expect a population of SNRs/SNe to contribute a cumulative spectral luminosity of $L_{\rm 5 GHz} = 1.4 \times 10^{20}$ W Hz$^{-1}$. The sum of these values ($L_{\rm 5 GHz} = 3.2 \times 10^{20}$ W Hz$^{-1}$) is only 8\% of the measured radio luminosity ($L_{\rm 5 GHz} = 3.9 \times 10^{21}$ W Hz$^{-1}$), strongly suggesting the radio emission is dominated by something other than H{\footnotesize{II}} regions and a population of SNRs/SNe.  We therefore favor an AGN origin for the radio emission in Mrk 709 S.    

If the X-ray source detected by \citet{reinesetal2014} is indeed associated with the massive BH (i.e., radio source 2), the X-ray luminosity would suggest the BH is radiating significantly below its Eddington luminosity.  For a BH mass of 10$^5~M_\odot$ and an X-ray to bolometric correction of $\sim 10$ \citep{vasudevanfabian2009}, the Eddington ratio would be $\sim 0.04$ (and even less for a more massive BH). A sub-Eddington BH such as this is expected to have a radiatively inefficient accretion flow (RIAF) and to be jet-dominated \citep[e.g.,][]{falckeetal2004}, which is consistent with the observed radio emission.  The optical line emission from such a weakly accreting BH could also easily be swamped by star formation and account for the H{\footnotesize{II}}-region-like line ratios measured in the SDSS spectrum obtained in a 3\arcsec\ aperture \citep{reinesetal2014}.

\subsection{Optical Evidence for an AGN -  [Fe X]$\lambda$~6374}
In addition to radio and X-ray evidence for an AGN, we also detect 
[Fe X]$\lambda$~6374 in the SDSS spectrum of Mrk 709 S. [Fe X] is a known AGN coronal line \citep{penston1984,Netzer2013}, and has a high-ionization potential of 262.1~eV \citep{Oetken1977}. This coronal line can be photoionized by the hard AGN continuum \citep[e.g., ][]{Nussbaumer1970,Korista1989,Oliva1994,Pier1995}, or more likely in Mrk 709 S, mechanically excited by out-flowing winds caused by radio jets \citep{wilson1999}.

\begin{figure}
    \centering
    \includegraphics[width=0.49\textwidth]{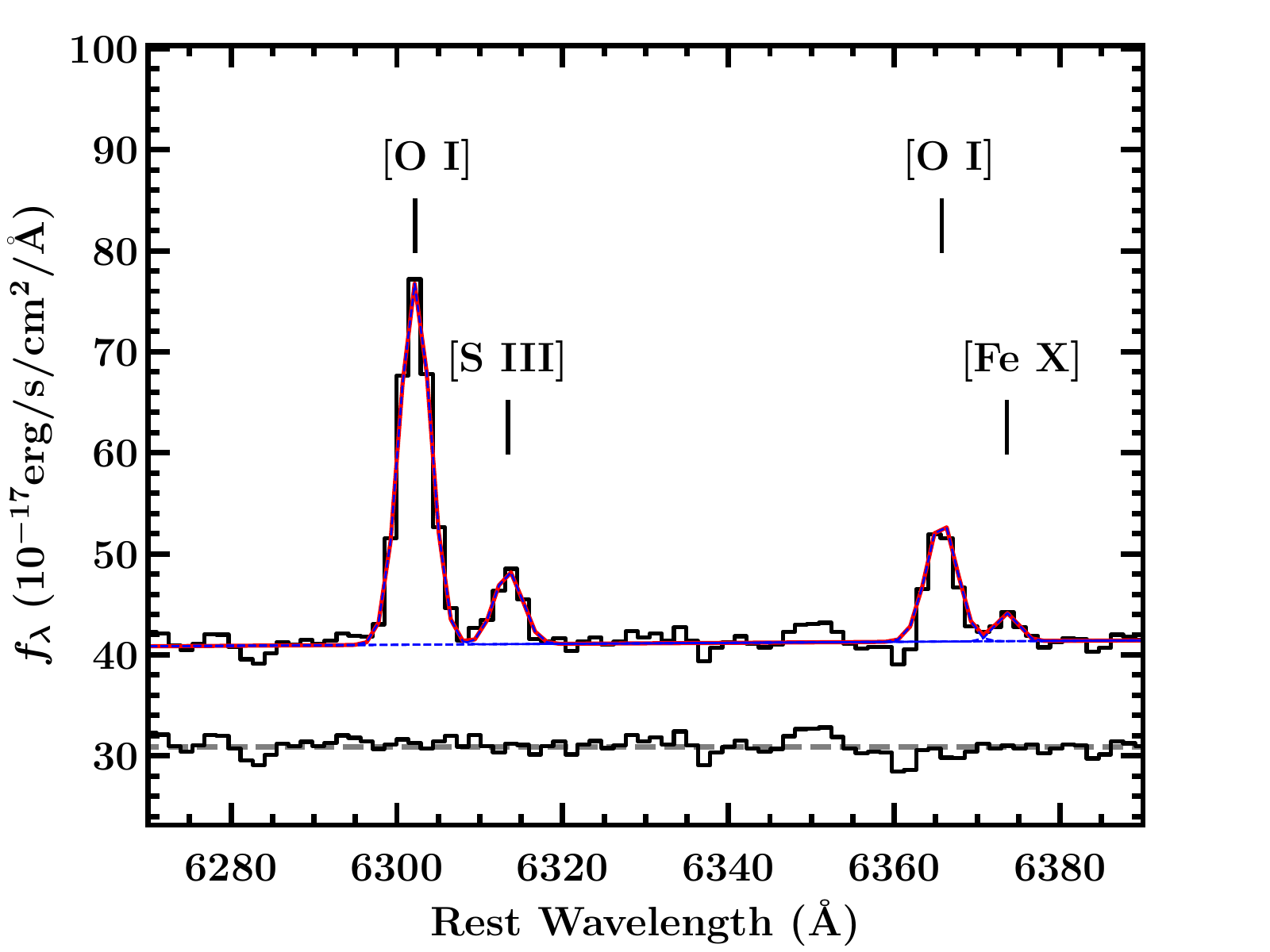}
\caption{The SDSS spectrum of Mrk 709 S, centered on the [O I] doublet. We label all four lines -- [O I]$\lambda\lambda$~6300,6363, [S III]$\lambda$~6312 and [Fe X]$\lambda$~6374 -- for reference. The data are shown in black, the total model in red and the individual components in blue. The residuals are shown below the fit. The AGN coronal line [Fe X] is detected with a $S/N=2.25$ with a peak that is $\approx3.3\sigma$ above the surrounding continuum.}
    \label{fig:fex}
\end{figure}

The fit to the [Fe X] emission line in the SDSS DR16 spectrum is shown in Figure~\ref{fig:fex}. The line has a flux of $9\times10^{-17}$~erg~s$^{-1}$~cm$^{-2}$, and a signal-to-noise ratio ${\rm S/N}=2.25$. The peak of the [Fe X] line is $\approx3.3\sigma$ above the surrounding continuum. While this is not a 3-$\sigma$ flux detection, the line is detected at the correct wavelength and could be diluted from host-galaxy light in the 3\arcsec\ SDSS aperture \citep{Moran2002}.  

Despite the presence of [Fe X], all three standard narrow-line diagnostic diagrams \citep{Kewley2006} indicate Mrk 709 S is powered by star formation. The lack of AGN-like line ratios can be explained by a combination of contamination from star formation in the host galaxy and a BH powered by a RIAF. For RIAFs with Eddington ratios of $\sim10^{-2}$ and lower, the broad lines, and potentially narrow lines, can disappear from the observed spectrum \citep{Trump2011}. Additionally, the relatively stronger radio emission in RIAFs \citep{Melendez2008} can drive relatively stronger radiative shock waves within the galaxy, which is likely causing the observed [Fe X] emission. Therefore, the X-ray, radio and optical emission are all best explained by a RIAF-driven active BH.

\section{Summary and Discussion}\label{sec:summary}

We have presented a study of the low-metallicity dwarf-dwarf galaxy merger Mrk 709 with the goal of investigating star formation and AGN activity. Our new observations include {\it HST} H$\alpha$ and continuum imaging, Keck spectroscopy, and VLBI with the High Sensitivity Array.  Our main results are summarized below.

\begin{enumerate}

    \item Our spectroscopy confirms that Mrk 709 S and Mrk 709 N are part of the same system and our {\it HST} imaging reveals a striking bridge of young massive star clusters between the two galaxies, indicating an interaction/merger.

    \item We estimate ages, ionizing luminosities and stellar masses for 26 H$\alpha$-selected (i.e., young) star clusters in the Mrk 709 system.  The youngest star clusters are in the bridge between the two dwarf galaxies, with ages $\lesssim 10$ Myr.  The clusters in the galaxies have ages of $\sim 10-15$ Myr.  Nearly all of the clusters we detect have stellar masses greater than $10^5~M_\odot$, which is the typical mass of a globular cluster.  
 
    \item Mrk 709 S is undergoing a clumpy mode of star formation resembling that in high redshift galaxies (see Section \ref{sec:masses} here; also \citealt{Elmegreen2009}). The clumps have stellar masses in the range $M_\star \sim  (1-8) \times 10^7~M_\odot$ and ionizing luminosities equivalent to $\sim 100 - 1400$ O-type stars.
     
    \item We present evidence confirming the presence of a massive BH in Mrk 709 S that was first identified by \citet{reinesetal2014} using VLA and {\it Chandra} observations.  In particular, we demonstrate that the radio luminosity is much too high to be produced by thermal HII regions and/or SNe/SNRs but can easily be explained by an AGN. The radio morphology at VLA resolution is suggestive of a core-jet object, which is consistent with our HSA non-detection.
    
    \item We also detect the AGN coronal [Fe X]$\lambda$~6374 emission line in the SDSS spectrum of Mrk 709 S, likely caused by shocks from the radio jets of the active BH. This result, in combination with the detected radio lobes, is consistent with a RIAF-powered low-luminosity AGN.

\end{enumerate}

We have presented definitive evidence that Mrk 709 N and S are in the midst of an interaction, resulting in a spectacular bridge of young massive star clusters between the two dwarf galaxies that resembles the ``beads on a string" mode of star formation \citep[e.g.,][]{elmegreen1983,mullanetal2011,tremblayetal2014}.  These clusters have high equivalent widths implying that they are extremely young ($\lesssim$ 10 Myr) and therefore must have formed in the bridge. Their birth can be attributed to the tidal forces from the interaction. Tidal features such as tails and bridges can result in a ``pile-up" of material that leads to dense cloud regions where it is possible for massive star clusters to form \citep{bournaud2010}.  There is also recent star formation in Mrk 709 N and S, likely due to an influx of new gas as the two components approach each other. Starbursts in merging galaxies most often occur at two points in the process of a merger: 1) when the galaxies make their first pass, and 2) at their final coalescence \citep{bournaud2010}. The existence of two distinct galaxies, as well as the tidal bridge, indicates that Mrk 709 is in the early stages of a dwarf-dwarf merger.  

There is also compelling evidence for an accreting massive BH in Mrk 709 S. Moreover, the BH has a position consistent with a $M_\star \sim 4 \times 10^7 M_\odot$ compact clump of star formation with a derived age of $\sim 13$ Myr. While the spatial overlap could be a projection effect, this is in stark contrast to the massive BH in the dwarf starburst galaxy Henize 2-10 \citep{Reines2011,reinesdeller2012,reinesetal2016}, which has no visible counterpart in optical or near-IR {\it HST} imaging.  However, there are several young massive star clusters in the vicinity of the BH in that galaxy with dynamical friction timescales $\lesssim 200$ Myr, suggesting Henize 2-10 may represent a rare snapshot of nuclear star cluster formation around a preexisting massive BH \citep{nguyenetal2014}. 
It is interesting to consider these different cases given that theories for the formation of BH seeds include collision in compact star clusters \citep{Loeb1994,Begelman2006,Lodato2006,Choi2015} as well as direct collapse scenarios \citep{Portegies2004,Devecchi2009,Davies2011,Lupi2014,Stone2017}.
 
Our findings demonstrate that Mrk 709, with a metallicity of only $Z \sim 0.1 Z_\odot$, may well be our best local analogue of high-redshift galaxies during the early stages of BH growth and globular cluster formation. Detailed multi-wavelength studies of additional BH-hosting star-forming dwarf galaxies are necessary to gain a more complete picture of the early stages of galaxy and BH evolution.




\acknowledgements
We thank the anonymous referee for their thoughtful comments and suggestions.
EMK acknowledges support for this project from the Montana Space Grant Consortium. Support for Program number HST-GO-14047.005-A was provided by NASA through a grant from the Space Telescope Science Institute, which is operated by the Association of Universities for Research in Astronomy, Incorporated, under NASA contract NAS5-26555. AER also acknowledges support for this paper provided by NASA through EPSCoR grant number 80NSSC20M0231. The National Radio Astronomy Observatory is a facility of the National Science Foundation operated under cooperative agreement by Associated Universities, Inc. The work of DS was carried out at the Jet Propulsion Laboratory, California Institute of Technology, under a contract with NASA.

Funding for the Sloan Digital Sky 
Survey IV has been provided by the 
Alfred P. Sloan Foundation, the U.S. 
Department of Energy Office of 
Science, and the Participating 
Institutions. 

SDSS-IV acknowledges support and 
resources from the Center for High 
Performance Computing  at the 
University of Utah. The SDSS 
website is www.sdss.org.

SDSS-IV is managed by the 
Astrophysical Research Consortium 
for the Participating Institutions 
of the SDSS Collaboration including 
the Brazilian Participation Group, 
the Carnegie Institution for Science, 
Carnegie Mellon University, Center for 
Astrophysics | Harvard \& 
Smithsonian, the Chilean Participation 
Group, the French Participation Group, 
Instituto de Astrof\'isica de 
Canarias, The Johns Hopkins 
University, Kavli Institute for the 
Physics and Mathematics of the 
Universe (IPMU) / University of 
Tokyo, the Korean Participation Group, 
Lawrence Berkeley National Laboratory, 
Leibniz Institut f\"ur Astrophysik 
Potsdam (AIP),  Max-Planck-Institut 
f\"ur Astronomie (MPIA Heidelberg), 
Max-Planck-Institut f\"ur 
Astrophysik (MPA Garching), 
Max-Planck-Institut f\"ur 
Extraterrestrische Physik (MPE), 
National Astronomical Observatories of 
China, New Mexico State University, 
New York University, University of 
Notre Dame, Observat\'ario 
Nacional / MCTI, The Ohio State 
University, Pennsylvania State 
University, Shanghai 
Astronomical Observatory, United 
Kingdom Participation Group, 
Universidad Nacional Aut\'onoma 
de M\'exico, University of Arizona, 
University of Colorado Boulder, 
University of Oxford, University of 
Portsmouth, University of Utah, 
University of Virginia, University 
of Washington, University of 
Wisconsin, Vanderbilt University, 
and Yale University.

Some of The data presented herein were obtained at the W. M. Keck Observatory, which is operated as a scientific partnership among the California Institute of Technology, the University of California and the National Aeronautics and Space Administration. The Observatory was made possible by the generous financial support of the W. M. Keck Foundation.

\software{astropy \citep{astropy2013,astropy2018}, matplotlib \citep{matplotlib}, photutils \citep{photutils}, pyspeckit \citep{pyspeckit}}

\bibliography{biblio}

\end{document}